\documentclass[journal,onecolumn,draftcls,12pt]{IEEEtran}

\usepackage[cmex10]{amsmath}
\usepackage{graphicx}
\usepackage{amsfonts}
\usepackage{mathrsfs}
\usepackage{txfonts}
\usepackage{amssymb}   
\usepackage{mathrsfs}
\usepackage{algorithm}
\usepackage{algorithmic}
\usepackage{color}
\usepackage{cite}      
\usepackage{psfrag}    
\usepackage{subfigure} 
\usepackage{url}       
\usepackage{stfloats}  
\usepackage{enumerate}
\usepackage{multirow}
\usepackage{booktabs}
\usepackage{setspace}
\usepackage[table]{xcolor}
\usepackage{longtable}
\usepackage{threeparttable}
\usepackage{makecell}

\newcommand{\qed}{\nobreak \ifvmode \relax \else
      \ifdim\lastskip<1.5em \hskip-\lastskip
      \hskip1.5em plus0em minus0.5em \fi \nobreak
      \vrule height0.75em width0.5em depth0.25em\fi}

\begin{document}

\title{Software Defined Networking Enabled Wireless Network Virtualization: Challenges and Solutions}
\author{\small Ning Zhang,
               Peng Yang,
               Shan Zhang,
               Dajiang Chen,
               Weihua Zhuang,~\IEEEmembership{\small Fellow,~IEEE,}\\
               Ben Liang,~\IEEEmembership{\small Senior Member,~IEEE,}
               Xuemin~(Sherman)~Shen,~\IEEEmembership{\small Fellow,~IEEE}

\thanks{Ning Zhang, Shan Zhang, Weihua Zhuang, and  Xuemin (Sherman) Shen are with University of Waterloo, Canada; Peng Yang is with Huazhong University of Science and Technology, China; Dajing Chen is with University of Electronic Science and Technology of
China; and Ben Liang is with University of Toronto, Canada.}}

\maketitle
%

\IEEEpeerreviewmaketitle

\begin{abstract}
Next generation (5G) wireless networks are expected to support the massive data and accommodate a wide range of services/use cases with distinct requirements in a cost-effective, flexible, and agile manner. As a promising solution, wireless network virtualization (WNV), or network slicing, enables multiple virtual networks to share the common infrastructure on demand, and to be customized for different services/use cases. This article focuses on network-wide resource allocation for realizing WNV. Specifically, the motivations, the enabling platforms, and the benefits of WNV, are first reviewed. Then, resource allocation for WNV along with the technical challenges is discussed. Afterwards, a software defined networking (SDN) enabled resource allocation framework is proposed to facilitate WNV, including the key procedures and the corresponding modeling approaches. Furthermore, a case study is provided as an example of resource allocation in WNV. Finally, some open research topics essential to WNV are discussed.

\end{abstract}
\section{Introduction}
The wireless communication network has achieved laudable success in providing users with seamless connectivity. However, new challenges are arising with the ever-increasing connected devices, tremendous data traffic growth, as well as the blossoming of services/use cases~\cite{andrews2014will}. For instance, it is predicted that the connected devices will reach 50 billion by 2020, while the traffic will continuously skyrocket, resulting in a 1000-fold increase, compared with that in 2010~\cite{zhang2015cloud}. In addition, a wide range of services and use cases will emerge with diverse service requirements in terms of data rate, reliability, latency, security, etc. Next Generation Mobile Network (NGMN) alliance envisions 25 representative use cases, which are grouped into eight families, such as Higher User Mobility, Massive Internet of Things (IoT), Extreme Real-Time Communications, and Ultra-reliable Communications \cite{osseiran2014scenarios}. 

The next generation (5G) wireless networks are expected to effectively accommodate the massive connectivity, the traffic surge, and the great diversity in services/use cases. Consequently, to boost network capacity, mobile network operators (MNOs) need to densely deploy network infrastructure such as small cell base stations, and add more radio spectrum \cite{zhang2016beyond}. As a matter of fact, MNOs have long suffered from the experience of over-provisioning to deal with peak demands, which results in very low resource utilization and extremely high capital expenditure (CAPEX). Resource sharing among MNOs is a potential way to reduce cost and improve resource utilization. For instance, 3GPP has specified two types of sharing with respect to the elements in the core network and the radio access network (RAN) \cite{3gpp2015}. On the other hand, the variety of services/use cases should also be supported efficiently. The current one-type-fits-all networks cannot achieve this goal due to the poor scalability, limited adaptability, and inflexibility. Those facts translate into the pressing need for a new solution to support tremendous traffic and a multitude of services/use cases in a cost-effective, flexible, and agile manner.

Wireless network virtualization (WNV) or network slicing\footnote{The phrases ``wireless network virtualization" and ``network slicing" are used interchangeably throughout this article.} has a great potential to meet the above needs efficiently. It slices the wireless physical infrastructure\footnote{Heterogenous resources from different sources can be put in a resource pool for sharing.} to create multiple virtual networks sharing the common resource pool \cite{kokku2012nvs,akyildiz2015wireless,liang2015wireless}. A slice or virtual network is a combination of network resources required to support a service/use case. The virtual networks are logically isolated from each other and any changes in a virtual network will not affect the operation of others. Furthermore, a virtual network can be customized according to the specific requirement of a service/use case, in terms of network topology, control policy, security mechanisms, etc. With WNV, the following benefits can be achieved: i) resource utilization can be greatly improved through efficient sharing by multiple virtual networks; ii) use cases or services with diverse requirements can be better supported since the virtual networks can be customized; and iii) it is cost-effective and flexible to roll out new services\footnote{A new service or technique can be introduced and tested in a virtual network, without affecting the running services in other virtual networks}.

From a technical perspective, WNV partitions and assigns a set of resources to different virtual networks, where the resources can be used in an isolated, disjunctive, or shared manner \cite{ONF2016}. Therefore, the essence of WNV is the network-wide resource allocation for different virtual networks. However, it is of great challenge considering the following facts: i) network-wide resource allocation should satisfy the end-to-end service requirements of virtual networks on the common infrastructure resource pool; ii) the traffic and network status are highly dynamic; and iii) the physical infrastructure exhibits multi-dimensional heterogeneity in devices spanning from access to core segments, and in network resources (e.g., bandwidth, computing, and storage).

In this article, we mainly focus on resource allocation for WNV to address the aforementioned challenges. A software defined networking (SDN) enabled resource allocation framework is proposed to support multiple virtual networks flexibly and dynamically. The procedure for resource allocation and the corresponding modeling approaches are provided. Specifically, based on the service description, the resource requirements are determined for each virtual network such that their Quality of Service (QoS) can be guaranteed. Then, based on the requirements from virtual networks and the physical resource constraints, sets of resources are despatched to different virtual networks. The resource despatched can be dynamically adjusted to adapt to the variations in the service requirement and in the network status. A case study is provided to illustrate resource allocation for WNV. Finally, some open research topics are discussed.

The remainder of this article is organized as follows. The motivations and enabling platforms of WNV are presented in Section~\ref{sec.Evolution}. The resource allocation for WNV and the technical challenges are discussed in Section~\ref{sec.Challenges}. The resource allocation framework for WNV is presented in Section~\ref{sec.Cloud}, with a case study in Section~\ref{sec.Study}. Some open research topics are discussed in Section~\ref{sec.Research}, followed by the conclusion in Section~\ref{sec.Conclusion}.

\section{Wireless Network Virtualization: Motivations and Conceptions}\label{sec.Evolution}
\subsection{Motivations}
Traditionally, MNOs invest and operate all the network infrastructure on their own. To scale up, this business model incurs heavy costs and increases operational complexity, leading to unsustainable and inflexible operation. In the upcoming 5G era, this model will be severely challenged due to the tremendous connectivity and massive data. MNOs has been constantly investing to upgrade the network and increase network capacity, while the profit almost keeps stagnant. That is mainly because the network operates as a pipeline. Moreover, 5G networks are expected to accommodate a great diversity in services/use cases efficiently. Those services/use cases have disparate requirements in terms of latency, reliability, data rate, etc. Traditional methods for augmenting network capacity and service provisioning cannot work well. A cost-effective, flexible, and agile solution is required to meet the needs of users and provide value-added service. 

\subsection{What is WNV} 
\begin{figure}[t]
    \centering
    \includegraphics[width=0.7\textwidth]{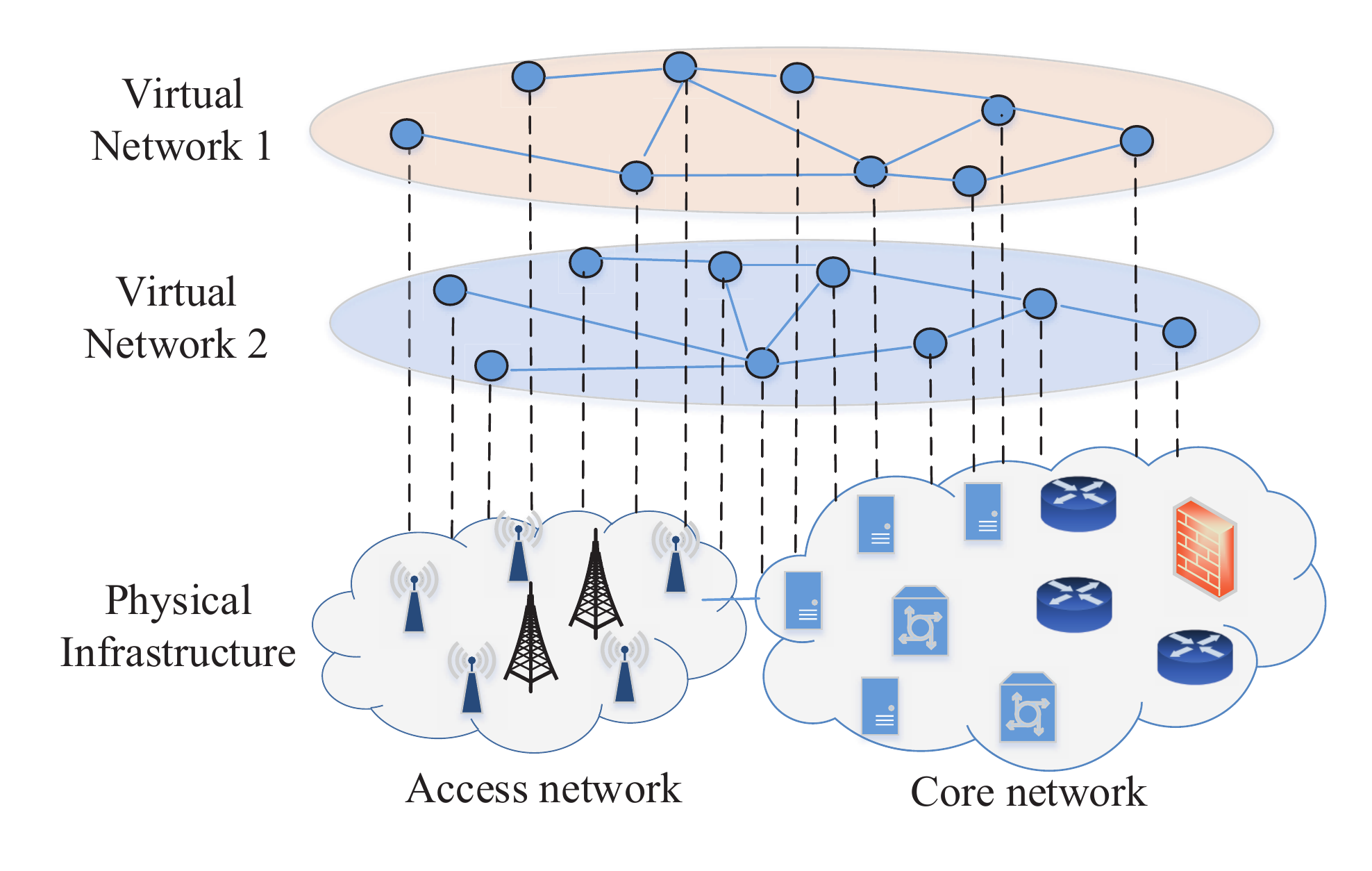}
    \caption{Wireless network virtualization.}
    \label{fig:slicing}
\end{figure}
WNV or network slicing can create multiple virtual networks sharing the same physical infrastructure, which helps lower the CAPEX. Furthermore, it can provide value-added service (e.g., provide Networking as a Service), by enabling virtual network customization for different applications or use cases. As shown in Fig.~\ref{fig:slicing}, WNV enables multiple logical, self-contained virtual networks on the common infrastructure spanning access and core segments. Different sets of resources are provisioned to create virtual networks on demand for services/use case driven networking to their respective users. Moreover, the physical resources in access and core network can be further virtualized, e.g., through network function virtualization (NFV) \cite{liu2016reliability}. With NFV, appropriate network functions can be efficiently chained such that service-oriented functions/policies can enforce on the traffic from a specific service/application, giving rise to service function chain (SFC) \cite{mehraghdam2014specifying}. The resulting SFC can be considered as a virtual network, with virtual functions chained in an appropriate way.


With WNV, a virtual network can be customized for particular services/use cases. Different from the virtual private networks or virtual local area networks, which mainly focus on a particular protocol layer, WNV allows virtual networks to be completely customized. Virtual network customization can be performed on the network topology, management policies, security mechanisms, resource management strategies, protocol stack, etc. To create a virtual network, the following three main steps can be performed. Firstly, the virtual network topology is customized, which consists of virtual nodes and virtual links and their corresponding capacities, based on the service description (e.g., user distribution, service request statistic, and end-to-end QoS requirements). Then, the virtual network is mapped to physical substrate network, which is also referred to as network embedding/mapping. Lastly, the service-oriented protocol stack can be tailored to better fit the services/use cases\cite{chiang2007layering}. Additionally, sub-virtual networks can be created on a virtual network to support multiple services simultaneously.

\subsection{Advantages}
WNV can achieve many benefits, as follows:

\subsubsection{Resource Sharing}
With WNV, the physical infrastructure resources can be shared by multiple virtual networks. Collections of resources are dispatched to multiple virtual networks on demand, helping avoid high CAPEX caused by the traditional over-provisioning method. In addition, it has a potential to significantly improve resource utilization.
\subsubsection{QoS Provisioning}
Through WNV, virtual networks can be customized for specific services/use cases. Therefore, WNV can better accommodate different use cases with diverse service requirements, compared with the conventional one-type-fits-all networking approach.
\subsubsection{Flexible Management}
Since virtual networks are isolated from each other, each virtual network can have its own control and management policies  independently. Any change in a virtual network will not disturb the services in other virtual networks.
\subsubsection{Service Innovation}
With WNV, an isolated virtual network can be created to deploy new services or techniques, which avoids disturbance to other running services. New techniques/products can be verified in production networks before widespread deployment.
\section{Wireless Network Virtualization: Essence and Enablers} \label{sec.Challenges}
\subsection{Essence of WNV}
The essence of WNV is network-wide resource allocation to support disparate service requirements of virtual networks. Specifically, a collection of resources is allocated by the infrastructure provider (InP) to constitute a slice (or a virtual network), spanning the access network and core network. For conventional resource allocation, resources are directly allocated to users from the operators. In contrast, in the context of WNV, groups of resources are firstly allocated to different virtual networks to satisfy the service requirements, considering their user distribution, service request statistics, average end-to-end service requirements, etc. Then, within each virtual network (VN), based on the instantaneous user requests (service request instances), resources of a virtual network are allocated to users to meet their requests. The former and the latter correspond to inter-VN and intra-VN resource allocation, respectively. Intra-VN resource allocation is similar to the conventional resource allocation, and this article mainly focuses on inter-VN resource allocation. Different from conventional resource allocation, inter-VN resource allocation is performed in a network-wide manner for multiple virtual networks to satisfy their end-to-end service requirements. Moreover, the resource allocation among virtual networks should be dynamically adjusted according to the spatial and temporal variations in traffic demands\footnote{Although static slicing can guarantee isolation, it results in low spectrum utilization.}.  

\subsection{Enabling Schemes}
\begin{figure*}[t!]
    \centering
    \includegraphics[width=\textwidth]{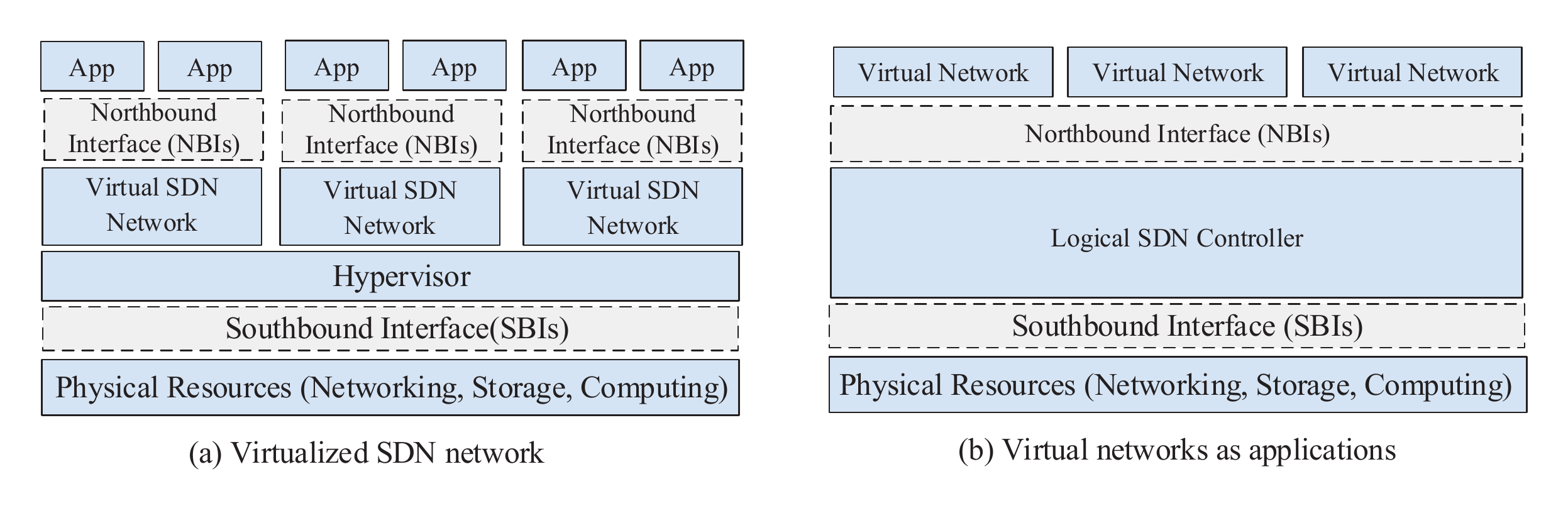}
    \caption{Wireless network virtualization.}
    \label{fig:ways}
\end{figure*}

To perform network-wide resource allocation for multiple virtual networks, the enabling schemes can be mainly classified into two categories: scheduling based \cite{kokku2012nvs,gudipati2014radiovisor} and software defined paradigm based \cite{ONF2016}\cite{blenk2016survey}. The former slices resources for different virtual networks by exploiting the existing schedulers or their extensions, e.g., the scheduling functions at base stations or routers. The latter relies on the new network paradigm of software defined networking (SDN) to allocate resources for multiple virtual networks, through the centralized control and open interfaces. Scheduling based approaches requires less modification to the existing system, which is fast to implement. However, they cannot fully support virtual network customization. Thanks to the features of SDN: i) separation of control plane and data plane; ii) logically centralized control; and iii) network programmability, MNV can be performed in a flexible and efficient way. Moreover, virtual network customization can be fully supported, where each virtual network can have its own SDN controller to enforce different control and management policies.

The SDN based schemes can be further divided into: hypervisor based \cite{blenk2016survey} and pure SDN controller based \cite{ONF2016}, as shown in Fig.~\ref{fig:ways}. For the former, an extra layer (i.e., hypervisor plane) is added into the system, above the southbound interface of the SDN layered architecture, as shown in Fig.~\ref{fig:ways}(a). A hypervisor introduced at network components helps create multiple virtual entities, e.g., multiple virtual base stations and gateways. A network level hypervisor coordinates local hypervisors to perform network-wide resource scheduling. For implementation, Flowvisor, and its extensions including VeRTIGO and OpenVirteX can be employed \cite{al2014openvirtex}. For the latter, with the global view of the network, the logically centralized SDN controller manages and allocates all the underlying physical resources for different virtual networks, as shown in Fig.~\ref{fig:ways}(b). Network virtualization can be regarded as an application on the top of the SDN control plane. Compared with the previous method, this approach requires the SDN controller to be more powerful to dynamically orchestrate virtual networks with diverse requirements. A comparison of those schemes is given in Table \ref{tab:comprison}.

\begin{table*}[t]\small
\caption{A comparison of different schemes.}\label{tab:comprison}
\centering
\begin{tabular}{|m{1.9cm}|m{4cm}|m{4.7cm}|m{4.5cm}|}
\hline
\multirow{2}*{Schemes}& \multirow{2}*{Scheduling based} & \multicolumn{2}{c|}{Software defined networking platform based}  \\ \cline{3-4}
             &       &  Hypervisor based  &  Pure SDN based  \\\hline
 Features        & Exploit existing schedulers or their extensions  & Rely on SDN platform; add an extra hypervisor layer   &  Rely on the three layers of SDN platform \\\hline
 Advantages       & Less modification to the existing system; easy to implement  & Fully support virtual network customization; simplified SDN controller &  Fully support virtual network customization \\\hline
 Disadvantages    & Limited customization; less flexibility  & Require new networking platform; more layers
    &  Require new networking platform; complex controller \\\hline
\end{tabular}
\end{table*}

\subsection{Challenges}
Resource allocation for WNV is of significance yet very challenging. The main challenges lie in the following aspects:
\begin{itemize}
\item Interference between adjacent nodes in different virtual networks, due to the broadcast channels;
\item Network-wide resource allocation to meet the requirement of virtual networks;
\item Support of multiple virtual networks with diverse end-to-end service requirements concurrently and efficiently;
\item Multi-dimensional heterogeneity in devices spanning from access to core networks, in radio access technologies (RAT), and in network resources (such as bandwidth, computing resource, and storage.);
\item Adaption to the changes in virtual network requests, variations in physical network status, and time-varying traffic distribution in virtual networks.
\end{itemize}


\section{Framework for WNV}\label{sec.Cloud}
\begin{figure*}[t!]
    \centering
    \includegraphics[width=16cm]{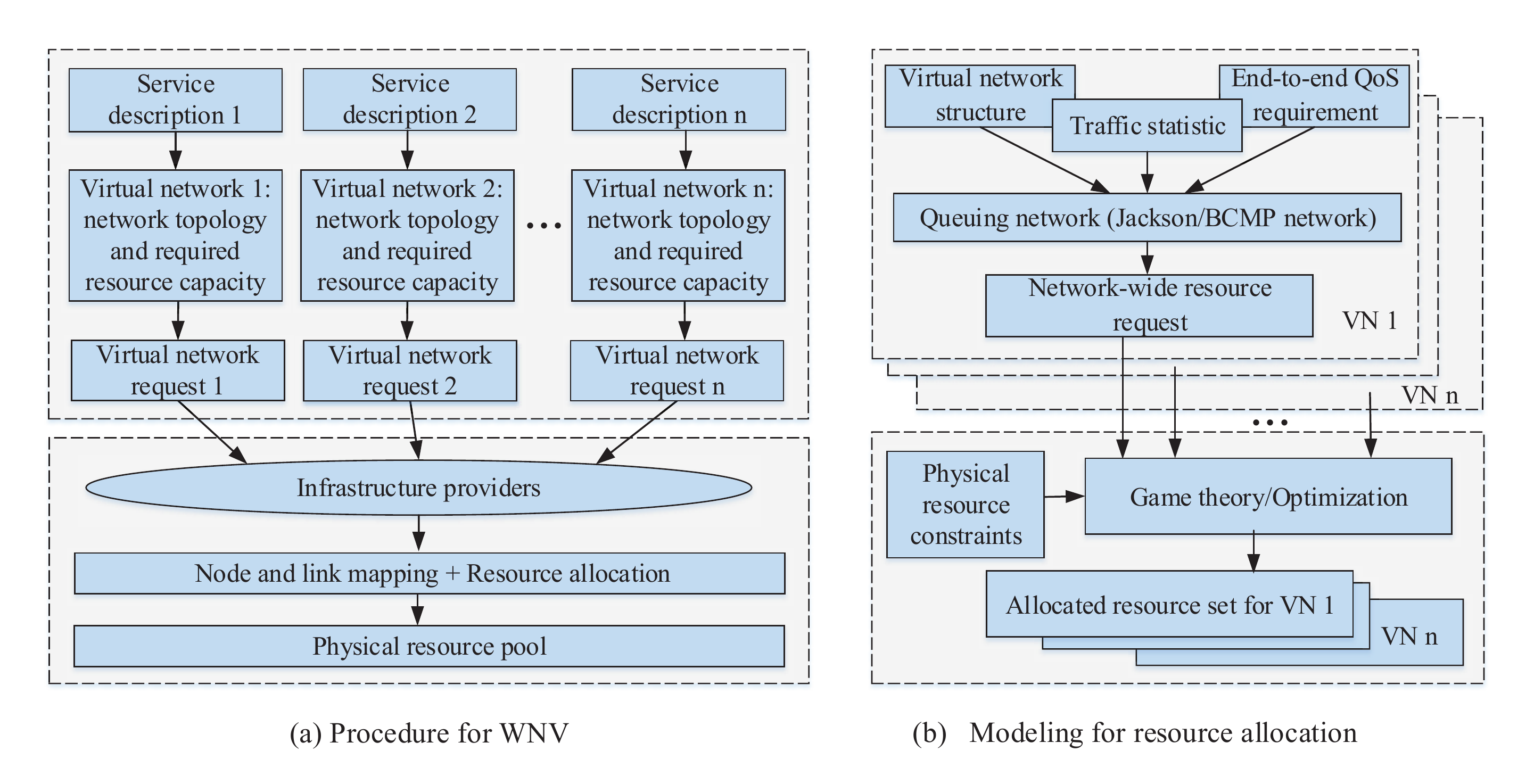}
    \caption{The resource allocation framework for WNV.}
    \label{fig:framework}
\end{figure*}

In this section, we propose a resource allocation framework for WNV, including the procedure, modeling, and approaches. It is illustrated in Fig.~\ref{fig:framework}.

\subsection{Procedure for WNV}
As shown in Fig.~\ref{fig:framework}(a), for each virtual network, a suitable virtual network topology is firstly established (in network planning) for the given service description. The service description may include users' distribution, statistics of service requests, and end-to-end service requirements. The virtual network topology determines the virtual nodes with different functions and their corresponding connections (i.e., virtual links). This process is similar to network planning in wireless networks, e.g., base station placement in the service area and core network design. Similarly, based on service requirements, SFC selects different virtual functions and connects them in an appropriate manner, which can also result in a virtual network. Based on the virtual network topology, the required capacity of virtual nodes and links are determined in order to satisfy the end-to-end service requirement (such as in terms of average delay and jitter). With the network-wide resource capacity requirements and the virtual network topologies, virtual network requests can be formed. Based on those requests from multiple virtual networks and the physical resource constraints, the virtual nodes are mapped to the physical nodes, the virtual links are mapped to the physical paths, and resources at individual physical entities are allocated for each virtual network\footnote{The ways for mapping include one-to-one, one-to-many, and many-to-many.}. 

%
%

\subsection{Resource Allocation Framework}
For the given virtual network topology and physical substrate network, resource allocation mainly includes two phases: network-wide resource request determination and resource dispatch for multiple virtual networks. The former determines the resources required for each virtual network to satisfy the QoS requirements, given the virtual network topology. Whereas the latter assigns resource sets to different virtual networks, based on the their requests and the physical resource constraints. To model the resource allocation procedure, potential approaches include optimization and game theory under different scenarios, as shown in Fig.~\ref{fig:framework} (b).

For the optimization based approach, the overall objective can be maximization of resource utilization defined by the number of virtual networks concurrently supported. In this case, each virtual network determines the resource capacity needed, and the InP maximizes the total virtual networks supported. Specifically, each virtual network firstly determines the optimal resource capacity required to satisfy the average end-to-end performance requirement. Taking delay requirement as an example, queuing networks such as the Jackson network or the Baskett-Chandy-Muntz-Palacios (BCMP) network can be employed to analyze the average end-to-end delay performance, given the virtual network topology, the traffic arrival rate, and the capacity of different network components. Then, optimization techniques can be employed to minimize the capacity required at different network components (i.e., the service rate $\mu_i$, where $i$ is the index of the entities) subject to the QoS requirement. With virtual network topology and optimal capacity requirement, virtual network resource requests are formed. Based on those requests, virtual networks are embedded into the physical substrate network through suitable mapping and resource allocation strategies, to maximize the number of concurrent virtual networks. The mapping problem may be formulated as integer programming and solved accordingly. The node and link mapping can be solved in an uncoordinated or coordinated fashion, considering the tradeoff between performance and complexity. Similarly, the mapping and resource allocation can also be solved separately or jointly.


When utilizing the resource from the InP, certain costs will be incurred for virtual networks. Therefore, virtual networks attempts to satisfy the QoS requirements with the minimal costs (e.g, leasing costs), while the InP aims to maximize its revenue from the virtual networks. In such a case, game theoretical approaches can be employed for the resource allocation procedure. In specific, based on queuing network, virtual networks determine the optimal resource capacity required to minimize the costs, i.e., $\sum_i p_i\mu_i$, where $p_i$ is the price for resource type $i$ set by the InP. For the InP, with those requests from different virtual networks, it maximizes the revenue by selecting suitable price vectors and mapping strategies. Contrary to the previous case, virtual networks have no fixed resource requests and can adjust the capacity requests to balance the achievable performance and the cost. Additionally, auction can also be considered, where the multiple virtual networks compete for the resource from the InP.

\subsection{Two-time Scale Operations}
Considering different characteristics of inter-VN and intra-VN operation, resource allocation for WNV is carried out in a two time-scale fashion. In small time scale, intra-VN resource allocation is dynamically performed to adapt to the instantaneous user requests, while inter-VN resource allocation remains unchanged. In large time scale, inter-VN resource allocation is adjusted dynamically. Dynamic inter-VN resource allocation is necessary because of: i) service requirement changes in existing virtual networks, due to the variation in traffic demand and service requests (e.g., various load statistics in different hours); and ii) new virtual network requests. For the first case, the resource allocation for different virtual networks will be adjusted by the InP, in order to adapt to the time-varying traffic statistics; For the second case, the InP not only adjusts resource allocation among existing virtual networks but also allocates resources for the newly coming virtual networks. An admission control policy is required such that the new virtual networks can only be admitted when both their requests and the service requirements of the existing virtual networks are satisfied.

%

\section{A Case Study}\label{sec.Study}
In this section, a case study is presented to demonstrate how virtual networks determine the resources required to support the respective services. Specifically, we consider that an InP supports virtual networks with different applications and service level agreements (SLAs), and demonstrate how the virtual networks respond to their application traffic, the SLAs, and the price of each resource set by the InP.

\subsection{Network Setting}
Suppose that two virtual networks coexist, which contain the RAN and core network elements. We consider the following four-node topology: each virtual network consists of a radio access node, a Serving Gateway (S-GW) node, an administrative P-GW node, and a Packet Data Gateway (P-GW) node each with independent service rate $\mu_i$ (i=1,2,3,4), respectively. Assume data traffic coming to the radio access node follows a Poisson process with average arrival rate of $\lambda$, and the data packet size follows exponential distribution with mean 1M. The stream reaches the S-GW node after RAN processing. With probability $q$, the data packets are for administrative processing and leave the network through the administrative P-GW, whereas the rest leave through the service P-GW for Internet service. We consider the latency requirement (e.g., the time for a packet to passes through the RAN and core network) in the SLA, and present how to meet the latency criteria with minimum costs. Jackson network is employed to analyze the above problem. Jackson's theorem indicates that the numbers of queued packets $N_i$ in all nodes are independent, and each node behaves as if its arrival stream $\lambda_i$ were Poissonian. Furthermore, we have $\lambda_1=\lambda$, $\lambda_2=\lambda$, $\lambda_3=q\lambda$ and $\lambda_4=(1-q)\lambda$, respectively.

\begin{figure}[t!]
\centering
\subfigure[]{\label{Setting1}
\includegraphics[width=0.6\textwidth]{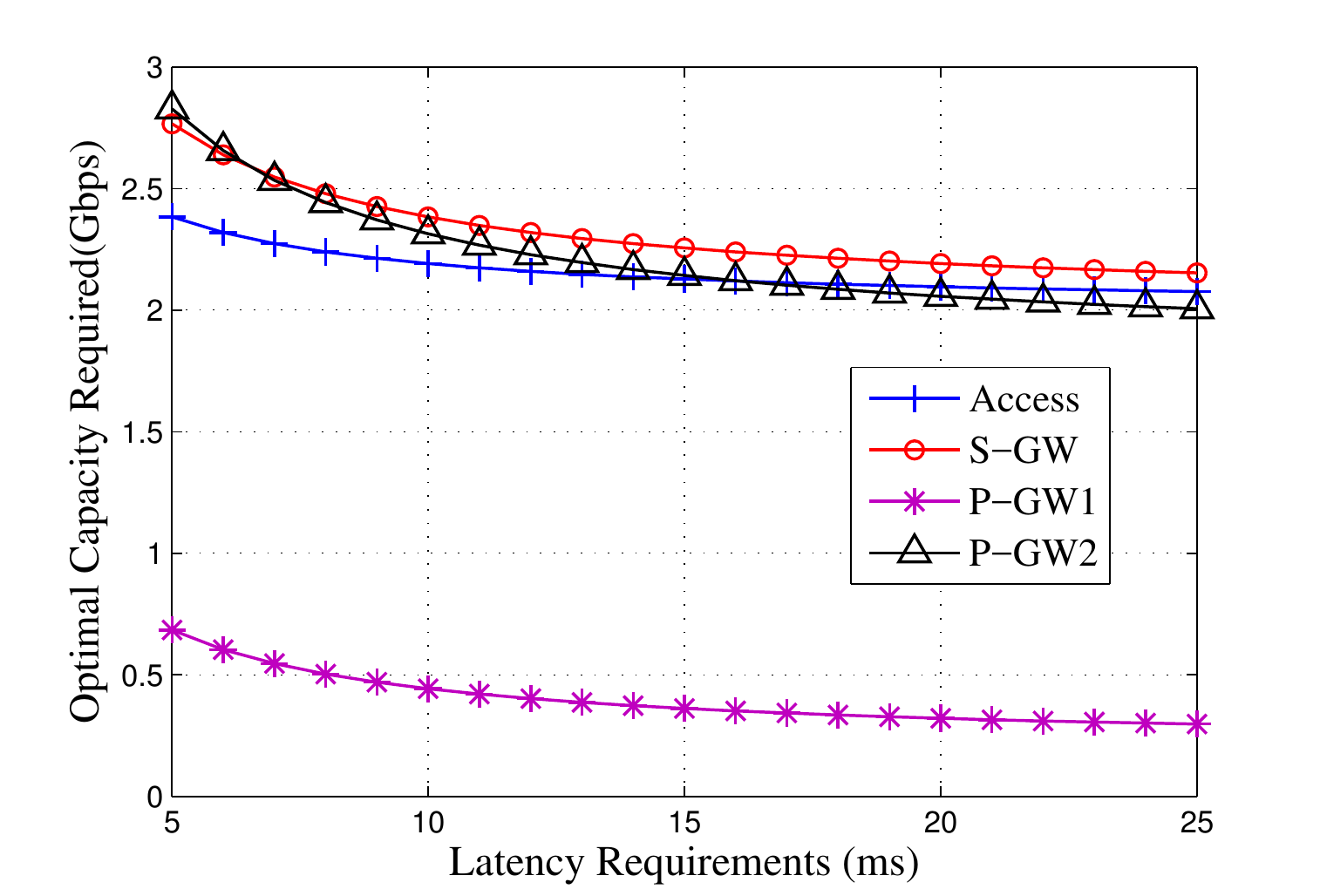}}
\subfigure[]{\label{Sheetting2}
\includegraphics[width=0.6\textwidth]{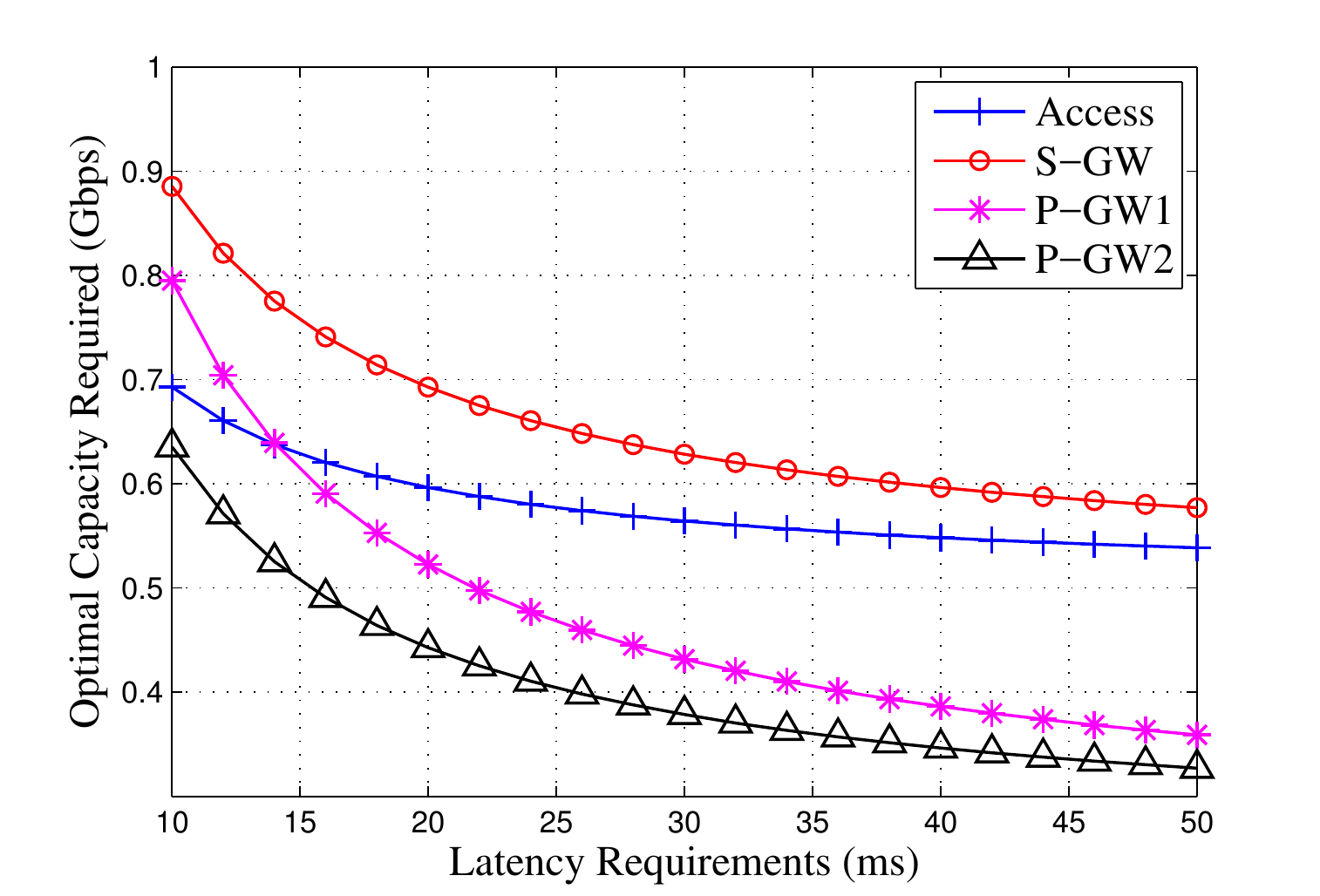}}
\caption{Optimal resource allocations of two applications versus latency requirement.}
\label{Casestudy}
\end{figure}
%
\subsection{Optimal Capacity Required}
If latency is required to be less than $T$ by the SLA, according to Jackson theorem and Little's law, we have the following constraint: $\sum_{i=1}^{4}\frac{\lambda_i}{\mu_i-\lambda_i}\leq\lambda T$. If the quoted unit price of service rates at each node is $p_i$, then the objective of each virtual network is to minimize the total cost $\sum_{i=1}^{4}p_i\mu_i$ under the preceding constraint. By applying the method of Karush-Kuhn-Tucker (KKT) multipliers, we can solve the optimization problem by minimizing $\sum_{i=1}^{4}p_i\mu_i+\alpha(\sum_{i=1}^{4}\frac{\lambda_i}{\mu_i-\lambda_i}-\lambda T)$. After solving this dual problem, the optimal processing capacity at each node can be obtained. 

We illustrate the optimal capacity under two different settings in Fig. \ref{Casestudy}. Suppose that the InP's price portfolio is $\textbf{p}=(p_1, p_2, p_3, p_4)=(0.8, 0.2, 0.05, 0.1)$. Fig. \ref{Setting1} shows the minimum resource required at each node under the scenario of high data volume uploading, such as live video streaming. The access arrival rate is 2000 Mbps, and 90\% of data are delivered to the service P-GW, while the remaining 10\% are delivered to the administrative P-GW for management function such as data statistics.  Fig. \ref{Sheetting2} shows the optimal resources required for low traffic applications such as network monitoring. The average data arrival rate is 50 Mbps, and 50\% of data are forwarded to the administrative P-GW. Fig. \ref{Casestudy} indicates that, when the latency requirement loosens to a certain level, the optimal service rate for each node stagnates at an amount slightly larger than that required to process traffic with intensity $\lambda_i$. Otherwise, a backlog may accumulate at that node and the latency will run out of control.

\begin{figure}[t!]
\centering
\includegraphics[width=0.6\textwidth]{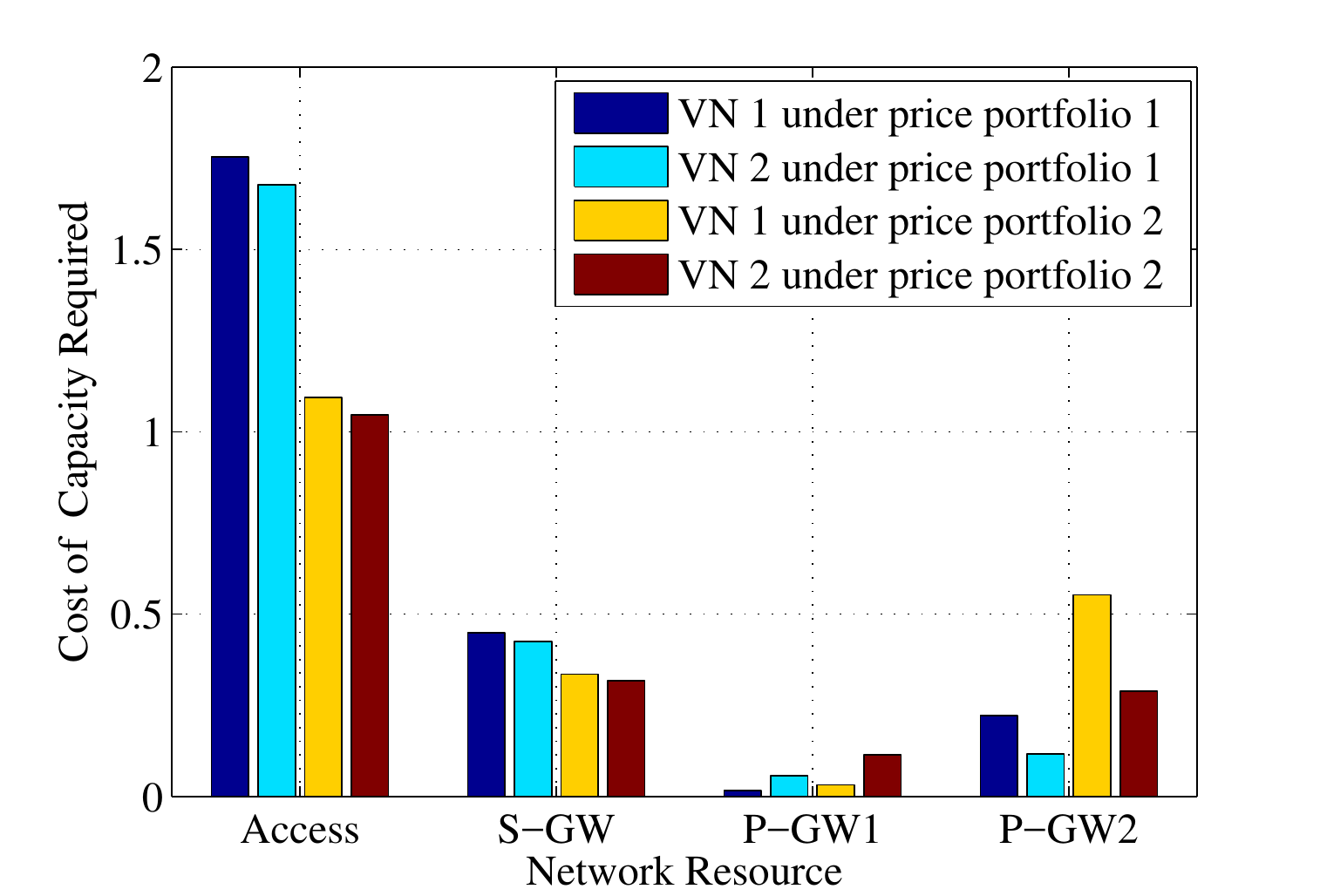}
\caption{VN's cost on each node under different InP pricing Strategies. Price portfolio 1: $\textbf{p}=(p_1, p_2, p_3, p_4)=(0.8, 0.2, 0.05, 0.1)$, Price portfolio 2: $\textbf{p}=(0.5, 0.15, 0.1, 0.15)$.}
\label{Cost}
\end{figure}

Fig. \ref{Cost} further shows the cost on each node with different price tags for the above two scenarios, where the corresponding latency requirements are $10$ ms and $20$ ms, respectively. With this framework, the optimal capacity required can be obtained for the considered case.


\section{Research Issues}\label{sec.Research}
In this section, some open research issues are discussed for the development of WNV.

%

\subsection{Automated Network Operation}
The SDN controller manages the network behaviors, based on the network state obtained from information acquisition and analysis. Because of the scale and volume of information, big data techniques such as data mining and machine learning can be adopted to extract knowledge and insights for guiding the decisions of SDN controllers. In order to achieve automated network operation to efficiently support WNV with minimal manual efforts, the interplay between SDN and big data needs further investigation.
\subsection{Dynamic and Fast Network Update}
After SDN controller makes decisions based on the current network state, the decision rules should be enforced in the network fast and efficiently. However, due to distinct capabilities and loads on the network entities, the decision rule update times can be various, resulting in inconsistency issue. Moreover, dependencies on the rules further make network update more complex. Dynamic and fast network update is of great importance to ensure the consistency and efficiency of WNV.
\subsection{Virtual Network Customization}
Virtual network can be customized to achieve service-oriented networking, taking into account the specific requirements of the services/use cases. Virtual network customization can be performed on virtual topology, functions of nodes, resource capacities, protocols, etc. Optimal virtual network topology and resource requirements need to be determined to satisfy the end-to-end service requirements. Moreover, the network protocol can also be customized, such as the protocol stack, function modules of a layer, and parameters. In this line, great research efforts are needed to enable full network customization.


\subsection{Joint Mapping and Resource Allocation}
With the requests from virtual networks, the InPs need to map the virtual nodes/links to the physical nodes/paths, which is known to be a NP-hard problem. Moreover, the InPs also need to determine the network-wide resource allocation for virtual networks. Jointly performing mapping and resource allocation will bring more gains, but it is more complex and challenging. An efficient scheme that jointly performs mapping and resource allocation is needed.


%

\subsection{Network Security}
Along with the benefits, WNV also poses various security issues. Since the SDN controller is responsible for dispatching and managing resources to different VNOs, it is vulnerable to denial-of-service (DoS) attacks, which can paralyze all the virtual networks operation. In addition, the SDN controllers can be compromised, whereby the attackers can disturb the normal operation of virtual networks, e.g., attackers intentionally change the resource allocation policy so that the experience of users in certain VNOs can be severely degraded. Therefore, the security issues be well addressed before the widespread deployment of WNV. 

\section{Conclusion}\label{sec.Conclusion}
In this article, we have proposed a resource allocation framework to pave the road towards wireless network virtualization. The procedure for WNV and the corresponding modeling approaches introduced provide some useful guidance and insights to study and realize WNV. Based on the proposed framework, many research topics can be facilitated, such as interaction among different parties and dynamic network-wide resource allocation. To accelerate the pace of WNV development and to better support diverse services/use cases in the 5G era and beyond, great research efforts on WNV are expected.

\bibliographystyle{IEEEtran}
\bibliography{SDN-WNV2.bbl}

\end{document}